# Disentangling the Evolution of Electrons and Holes in photoexcited ZnO nanoparticles

Christopher J. Milne[1,2]*, Natalia Nagornova[3], Thomas Pope[4], Hui-Yuan Chen[3], Thomas Rossi[3], Jakub Szlachetko[2,5], Wojciech Gawelda[1,5,6], Alexander Britz[1,7], Tim B. van Driel[8], Leonardo Sala[2], Simon Ebner[2], Tetsuo Katayama[9,10], Stephen H. Southworth[11], Gilles Doumy[11], Anne Marie March[11], C. Stefan Lehmann[11,12], Melanie Mucke[13], Denys Iablonskyi[14], Yoshiaki Kumagai[14], Gregor Knopp[2], Koji Motomura[14], Tadashi Togashi[9], Shigeki Owada[10], Makina Yabashi[10], Martin M. Nielsen[8], Marek Pajek[5], Kiyoshi Ueda[10,14], Rafael Abela[2], Thomas J. Penfold[4] and Majed Chergui[3]*

1 European XFEL, D-22761 Hamburg, Germany
2 SwissFEL, Paul Scherrer Institute, 5232 Villigen-PSI, Switzerland
3 Lausanne Centre for Ultrafast Science (LACUS), ISIC, FSB, Ecole Polytechnique Fédérale de Lausanne, CH-1015 Lausanne, Switzerland
4 Chemistry – School of Natural and Environmental Sciences, Newcastle University, Newcastle upon Tyne, NE1 7RU, UK
5 SOLARIS National Synchrotron Radiation Centre, Jagellonian University, 30-387 Krakow, Poland
6 IMDEA Nanoscience Institute, Calle Faraday 9, Campus Cantoblanco, 28049 Madrid, SPAIN
7 The Hamburg Centre for Ultrafast Imaging, Luruper Chaussee 149, 22761 Hamburg, Germany
8 Department of Physics, Technical University of Denmark, 2800 Kongens Lyngby, Denmark
9 Japan Synchrotron Radiation Research Institute (JASRI), Kouto 1-1-1, Sayo, Hyogo 679-5198, Japan
10 RIKEN SPring-8 Center, Kouto 1-1-1, Sayo, Hyogo 679-5148, Japan
11 Argonne National Laboratory, 9700 S. Cass Ave., Argonne, Illinois 60439, USA
12 Advanced Research Center for Nanolithography (ARCNL), Science Park 106, 1098 XG Amsterdam, Netherlands
13 Department of Physics and Astronomy, Uppsala University, 751 20 Uppsala, Sweden
14 Institute of Multidisciplinary Research for Advanced Materials, Tohoku University, Sendai 980-8577, Japan

*Corresponding authors: christopher.milne@xfel.eu; majed.chergui@epfl.ch*

**Abstract:**

The evolution of charge carriers in photoexcited room temperature ZnO nanoparticles in solution is investigated using ultrafast ultraviolet photoluminescence spectroscopy, ultrafast Zn K-edge absorption spectroscopy and *ab-initio* molecular dynamics (MD) simulations. The photoluminescence is excited at 4.66 eV, well above the band edge, and shows that electron cooling in the conduction band and exciton formation occur in <500 fs, in excellent agreement with theoretical predictions. The X-ray absorption measurements, obtained upon excitation close to the band edge at 3.49 eV, are sensitive to the migration and trapping of holes. They reveal that the 2 ps transient largely reproduces the previously reported transient obtained at 100 ps time delay in synchrotron studies. In addition, the X-ray absorption signal is found to rise in ~1.4 ps, which we attribute to the diffusion of holes through the lattice prior to their trapping at singly-charged oxygen vacancies. Indeed, the MD simulations show that impulsive trapping of holes induces an ultrafast expansion of the cage of Zn atoms in <200 fs, followed by an oscillatory response at a frequency of ~100 cm$^{-1}$, which corresponds to a phonon mode of the system involving the Zn sub-lattice.

# I. Introduction :

Transition metal oxides (TMO), such as Titanium dioxide ($TiO_2$) and Zinc oxide (ZnO), are large-gap (>3.2 eV) semiconductors that have been attracting considerable interest in the past three decades or so, due to their remarkable optical properties, robustness under ambient conditions, abundance, and ease of preparation. [1,2] This makes them potential candidates for photovoltaic and photocatalytic applications, [3–7] detectors for high energy radiation, [8] transparent conductive oxides, [9] lasing, [10] pressure sensors with optical read-out, [11,12] etc. Their large band-gaps also offer the advantages of higher breakdown voltages, the ability to sustain large electric fields, lower noise generation, and high temperature and high-power operation.

These current and potential applications rely on the generation of charges and their subsequent evolution via electron-electron and electron-phonon scattering, diffusion through the lattice, thermalization and eventually, localization either as self-trapped excitons (intrinsic trapping by electron-phonon coupling) or at defects (extrinsic trapping), followed by radiative and/or non-radiative electron-hole recombination. The initial events following photoexcitation take place at ultrashort time scales and need to be described in detail in order to reach optimal performances of the material for a specific application. This requires tools that can probe the evolution of charge carriers in real-time, are specific to both the valence (holes) and conduction (electrons) bands and are, ideally, element-selective.

In the past 25 years or so, a large variety of ultrafast optical methods have been used to monitor the charge carrier dynamics in TMO's. In these experiments a non-equilibrium distribution of electrons and holes is created upon above band-gap excitation and the ultrafast (femtoseconds to picoseconds) evolution of charge carriers is monitored using different probes from the Terahertz (THz) to the ultraviolet (UV) and visible spectral range. [13–22] These probes are generally tuned to the intra-band transitions and therefore monitor the free carrier response, which does not always distinguish between the electron and hole responses, nor does it provide an unambiguous identification of trapping. Deep-UV probing of the inter-band transitions has also been implemented, as it can in principle distinguish between the dynamics occurring in the valence band (VB) and the conduction band (CB). [22–27] However, the TA signal in this case is sensitive to the joint density-of-states (DOS) of the two bands, and therefore, when the evolution of free carriers is on comparable timescales, they are also difficult to disentangle. Furthermore, charge carrier localization at defect states cannot be unambiguously determined. To solve the latter, ultrafast sum-frequency generation with a white light continuum resonant with the in-gap defect states was implemented, [28] reporting sub-picosecond cooling times of the electron in the CB.

Photoluminescence (PL) is sensitive to the DOS in the CB. In the ultrashort time domain, it can selectively detect the cooling of electrons down to the bottom of the band, as well as formation of free excitons. [27,29] However, to our knowledge, nearly all time-resolved PL studies of TMOs have focussed on the electron-hole recombination (see table S1 of ref. [30] and ref. [28]), which is on the tens of ps to ns time scales. This is, in particular, the case for the system of interest here, Zinc oxide (ZnO). In this work, we implemented ultrafast PL up-conversion spectroscopy in the UV in order to monitor the relaxation of electrons in the CB via the rise of the excitonic emission close to the band-gap (BG).

We complement the ultrafast PL study by ultrafast hard X-ray absorption spectroscopy (XAS). Over the past decade, time-resolved soft and hard XAS has increasingly been used to investigate the fate of charge carriers in photoexcited TMOs and perovskite nanoparticles (NPs) in colloidal solutions. [30–37] In TMOs, the oxygen 2p-orbitals dominate the VB, while the metal 3d-orbitals dominate the CB. [38] Therefore, the element-specificity of XAS implies to a certain extent, an electronic band selectivity, as was nicely illustrated in ref. [37]. In the case of $TiO_2$, [31–34,37] the ps and fs hard XAS studies showed signals that were predominantly due to changes of the Ti oxidation state from 4+ to 3+ and are therefore mostly sensitive to electron trapping. Ultrafast soft XAS at the O K-edge and the Ti $L_{2,3}$-edges could simultaneously monitor both the hole and the electron trapping in photoexcited anatase $TiO_2$ nanoparticles and single crystals. [37]

ZnO is a direct band-gap (3.4 eV) semiconductor that has a bulk exciton binding energy of 60 meV at room temperature, native n-type doping and high conductivity, [1] conferring to this material a high potential for optoelectronic applications in the visible and UV photon energy range. The band structure of ZnO exhibits a splitting of the top-most VB into three sub-bands usually termed as A, B, and C, due to a combination of crystal field and spin–orbit coupling. [39,40] Transitions between these bands and the CB dominate the optical absorption spectrum (Figure S1) at different polarizations. However, the band edge absorption stems from the vicinity of the VB maximum. [41,42]

Different to $TiO_2$ where the electronic configuration of the metal atoms is $d^0$, in ZnO it is $d^{10}$ and therefore, the metal atom cannot be reduced. Nevertheless, in a recent study of ZnO

NPs photoexcited at 355 nm using time-resolved Zn K-edge XAS and X-ray emission spectroscopy (XES) with 80 ps resolution, [30] dramatic changes were observed in the X-ray near-edge structure (XANES) spectra and the extended X-ray absorption fine structure (EXAFS) spectra. The time-resolved Zn K$_\alpha$ and K$_\beta$ emission lines, which are sensitive to the electronic structure, showed however only weak charge density changes on the Zn atoms. This implied that the XANES and EXAFS spectral changes are largely due to structural effects. These spectral changes were rationalised by noting that ZnO is rich in singly-charged Oxygen vacancies ($O_{vac}^+$) [43] and that upon photoexcitation, the free hole charge carriers generated in the VB of the material, migrate and get trapped at the $O_{vac}^+$ defects to form doubly-charged oxygen vacancies ($O_{vac}^{++}$). Previously, theoretical calculations had shown that upon formation of a doubly-charged oxygen vacancy, [44,45] a dramatic increase of the $O_{vac} - Zn$ distance occurs, displacing four Zn atoms per trapped hole charge. From the XANES and EXAFS features, the estimated $O_{vac} - Zn$ bond length increase was found to be ~15% of its value prior to trapping. [30] This result demonstrated the ability to detect hole trapping in ZnO by means of hard X-ray Zn K-edge absorption spectroscopy, even if the Zn atoms are not subject to significant electronic structure changes.

In order to disentangle the evolution of the electrons and holes in photoexcited ZnO, here we combine femtosecond-resolved UV PL up-conversion [46] studies with femtosecond-resolved Zn K-edge XANES. The UV PL study was carried out at the Lausanne Centre for Ultrafast Science (LACUS) under 4.66 eV excitation, which is well above the optical BG of 3.4 eV (Figure S1). [1] The fs XANES measurements were carried out at the SACLA X-ray Free Electron Laser (XFEL) in Japan at an excitation energy of 3.49 eV, in order to minimize effects due to energetic electrons. Indeed, this energy is resonant with the blue wing of the first exciton and the red edge of the band gap absorption (Figure S1). In order to rationalise the X-ray results, we performed *ab-initio* molecular dynamics (MD) simulations of the structural rearrangement around the newly formed doubly-charged $O_{vac}$. Details of the experimental and theoretical procedures and set-ups are given in the SI.

Our results show that electron cooling in the CB and formation of the exciton occur in <500 fs, in very good agreement with theoretical predictions, [41,42] while the time scale for the hole response of ~1.4 ps is governed by hole diffusion through the lattice and its trapping as the ensuing structural response is prompt according to the MD simulations.

## II. Results:

*II.1 Femtosecond photoluminescence studies:*

The steady-state PL spectrum of ZnO NPs at room temperature (figure S2) consists of a band around 3.37 eV and a broad band centred at ~2.3 eV. [28,47] The former is due to an excitonic electron-hole recombination between the CB and VB, while the latter has been attributed to recombination of CB electrons with hole defects (Oxygen vacancies) that form trap states within the band gap. [28,30,47,48] The time-resolved PL studies of RT ZnO in various forms (crystals, films or nanoparticles) report different e-h recombination times for the excitonic and the green luminescence (Table S1 of ref. [30]), which in addition are sensitive to sample preparation. [28]

Figure 1 shows a two-dimensional time-energy plot of the ZnO NP's excitonic PL over the first 100 ps after fs-excitation at 4.66 eV, while the inset zooms into the first 5 ps. The low-energy part of the emission is cut at 3.2 eV because of contamination by the strong scatter of the remnant 400 nm light from the laser. The high-energy part of the spectrum extends out to 3.9 eV, however as the band-edge is at ~3.5 eV, this implies that any higher energy PL is in part reabsorbed by the sample. The remarkable feature here is that the PL is already at the energy of the exciton PL band from the earliest times, but the inset shows that the PL extends out to ~3.8 eV, i.e. well above the gap, at the earliest times. Cuts of the time-energy plot at different time delays up to 10 ps are shown in figure 2. It can be seen that the emission grows within the first few hundreds of fs, at almost the same energy as the steady-state excitonic PL. The dependence of the PL intensity as a function of pump fluence is linear, as shown in figures S3 to S5. In particular, the time traces recorded at the maximum (3.31 eV) of the PL are shown in figure S4 up to 5 ps for different fluences. Figure 3 shows the kinetic trace of the PL at the same energy for both long and short (insert) time windows. Figure S6 compares the time trace of the signal at early time with the Instrument response function (IRF), clearly showing that the signal's rise time is significantly longer. In figure 3, the long-time trace exhibits a biexponential behaviour and can be fitted with time constants of ~6.5 ps and ~40 ps (Table 1). The fit of the short time traces convoluted with the IRF (approximated as a Gaussian) yields a value of ~450 fs for the rise time, independent of the fluence (table S1). Considering that the early time PL appears almost resonant with the steady-state one, this suggests that the rise time of the PL integrates the cooling time of the electrons in the CB, as well as the formation of a relaxed excitonic state that yields the PL. With an excitation of 4.66 eV, i.e. 1.33 eV above the minimum of the CB, we can conclude that the electron cooling to the bottom of the CB occurs at a rate of approximately 3 meV/fs (1330 meV/450 fs). This is in excellent agreement with the predictions by Zhukov et al [41] that cooling of the high excess energy electrons is ultrafast, as they exploit the entire optical phonon phase space for energy dissipation.

The steady-state PL spectrum of ZnO exhibits a rich fine structure with several lines, separated by a few tens of meV's, attributed to different excitonic transitions, [1,28] and it is therefore a composite band. The decay times of ~6 ps and ~40 ps may be due to different transitions therein and /or relaxation processes within the manifold of states giving rise to the PL. [28] In table 1, we compare the times scales of the PL with those found using UV pump/UV continuum probe TA.[48] In the latter case, the pump energy was 295 nm (4.20 eV), close to the present 4.66 eV excitation, and the probe was a continuum spanning the 280-360 nm range. The excitonic band was found to be bleached at t=0 and its recovery timescales are given in table 1. While some of the time scales may bear a correspondence with the PL ones, it is difficult to be affirmative, as the TA is sensitive to the joint DOS of the VB and CB.

*II.2 Femtosecond Zn K-edge absorption spectroscopy*

Figure 4 shows the Zn K-edge XANES spectrum of ground state ZnO NPs (black trace) and the transient at 2 ps obtained upon 3.5 eV excitation, along with the transient spectrum previously obtained at 100 ps time delay. [30] It can be seen that most of the features of the latter are already present in the 2 ps transient, but with somewhat different relative amplitudes. This implies that the most significant signatures of hole trapping, discussed in ref. [30], are already present 2 ps after photoexcitation. Most of time-resolved XAS studies have focussed on the XANES region as it provides more contrasted signals. [50–53] However, one of the striking aspects of the ps-XAS study of ZnO [30] is that the amplitude of the transient EXAFS was of comparable amplitude as the transient XANES. Figure 5 compares the entire transient XAS (XANES and EXAFS) spectra at 2 ps and at 100 ps after band gap photoexcitation. It can be seen that the two transients are quite similar both in the XANES and in the EXAFS region. In ref. [53], the transient linear dichroism XANES spectroscopy of epitaxial ZnO nanorods on monocrystalline quartz substrates was reported at 100 ps time delay. The similarities between transient and the temperature-induced XANES and EXAFS spectra led the authors to conclude that thermal effects are predominant but also to extract the actual electronic effects from their transients. Nevertheless, while not fully ruling out a thermal effect in the present transient XANES and EXAFS spectra, we do not hold it for predominant on the basis of the following reasons: a) the samples in ref. [53] consisted of 1-2 µm-long nanorods with a diameter of ~70 nm. The nanorods did not have a fluid as an energy dissipation bath and as pointed out by the authors, the long decay times (200-300 ns) of their transients were due to heat diffusion along the rods. [53] Our samples consist of spherical NPs in solution, implying a more efficient heat dissipation; b) the fact that the early time XANES and EXAFS reported here show such a resemblance with the 100 ps transient is quite remarkable and speaks against a predominance of thermal effects; c) The band-edge in the UV absorption of ZnO is very sensitive to temperature, and it undergoes a significant red shift with increasing temperature, [54] however the ultrafast TA probing across the band gap shows an opposite trend, which again rules out a heating effect; [49] d) This is furthermore so that for the present X-ray measurements, the excess energy deposited to the system is minimised by the 3.49 eV excitation we used.

The temporal evolution of the signal at 9.67 keV, where the amplitude of the X-ray transient is largest, is shown in Figure 6 for early times, while Figure S7 show the kinetic traces at long and intermediate times. Figure 6 shows that the negative amplitude signal rises from zero and it reaches a plateau by ~5 ps, while the long-time trace (Figure S7) shows a recovery that could be mono- or biexponential. We could satisfactorily fit the short and long-time traces with a function consisting of a rising component and one decay component convoluted to the cross-correlation of the experiment approximated as a gaussian (see § S2). The fit is shown in Figures 6 and S7 and it yields a rising component of 1.4±0.1 ps and a recovery one of 126±44 ps. In the previous XAS study of ZnO NPs with 100 ps resolution, [30] the kinetic trace at the same energy was scanned to longer delay times and it exhibited a biexponential decay with time constants of 200±130 ps and ~1.2±0.3 ns (Table 1). The former is in the same scale as the recovery component reported here.

*II.3 Ab-initio Molecular Dynamics simulations*

The main focus of the present XAS measurement is the short component of ~1.4 ps, which may reflect either a structural response ensuing an ultrafast trapping of the holes, or the diffusion time of the holes in the lattice to the traps and their localization at the traps, followed by latter's ensuing structural relaxation. In order to verify these hypotheses, we carried out *ab-initio* molecular dynamics (MD) simulations. The details of the calculations are given in § S.3.

Figure 7 shows the percentage change ($\Delta d$) in the distance of the atoms from the oxygen vacancy upon switching the vacancy from singly- to doubly-charged, which mimics a prompt hole trapping. The red trace shows the first coordination shell, *i.e.* the average of the distance between the Zn atoms around the oxygen vacancy upon hole trapping. It undergoes a rapid increase up to just over 40 % in the first 200 fs, followed by an oscillatory behaviour with a period around 0.33 ps that damps away in >1.5 ps. The blue/black trace shows the response of the second/third coordination shell, *i.e.* the next-neighbour O/Zn shells around the vacancy. Clearly only the first shell around the vacancy responds, while the next neighbour shells do not, as is typical of an optical phonon. The value of the cage expansion converges to around 25%, which is close to original value of 23 % calculated by Janotti and De Walle, [45] while a value of 15% was reported from the transient XANES at 100 ps. [30] This deviation may be due to the fact that in the latter case, thermal effects had completely been neglected in the simulations of the transients. The oscillation period is quite short (about 327-345 fs) and it most likely correspond to the $E_2$ mode at 99 cm$^{-1}$ (336 fs) reported in the Raman spectrum of ZnO and attributed to motion of Zn atoms. [55] Its relatively long damping time is also in line with the narrow linewidth reported in the Raman spectrum. In addition, in PL spectra of low temperature ZnO, phonon replicas at this energy have been reported for the excitonic transition. [56] This shows that the doubly-charged oxygen vacancy has the characteristics of a small hole polaron.

In light of the above XAS and MD results, we discuss below the fate of holes in the system, after analysing the optical ultrafast PL. The time constants extracted from the present ultrafast UV PL and fs-XAS studies are collected in table 1 and are compared to those obtained in the previous ps-XAS [29] and UV probe TA [49] studies.

III.     Discussion:

The main focus of the present work are the early times of the evolution of charge carriers prior to the e-h radiative recombination. Regarding these times, a number of (mostly) UV-visible TA studies have been carried out on different types of ZnO: epitaxial thin films, [23] various nanostructures (dots, rods, wires and ribbons), [24,57] ZnO/ZnMgO multiple quantum wells, [25] molecular beam epitaxy films and single crystals. [28] These were generally carried out at RT and for different excitation fluences, and they concluded that the charge carrier relaxation spans timescales from 200 to 1000 fs. Ultrafast 2-photon photoemission studies of single crystals of ZnO excited at 4.19 eV were also carried out, reporting electron cooling times of 20-40 fs, followed by formation of a surface exciton on a time scale of ~200 fs. [58] Considering that these studies used quite similar excitation energies, well above the band gap, the fact that the values of the reported times are so scattered, has to do with the pump fluence, the sample morphology and possibly, the environment. [24,57]

Using density functional theory (DFT) calculations, Zhukov and co-workers [41,42] identified two regimes of electron-phonon cooling depending on the electron excess energy with respect to the highest phonon energy. At high electron excess energies, the whole phonon dispersion acts on the electron cooling, whose time scale spans from 100 to 500 fs, while below the cut-off of the highest energy phonon, only phonons with an energy lower than that of the electron excess energy will play a role in the cooling and the energy loss time can span a very large range from sub-100 fs up to 10 ps with a rapid increase below an excess energy of 20 meV due to the reduction of the available phonon phase-space. For the holes, [41,42] the energy loss time at any excess energy was found to be about three times smaller than the electron energy loss time.

In the PL experiment, $\tau_r$ in table 1 corresponds to electron cooling in the conduction band and formation of the exciton, in very good agreement with the above theoretical predictions. [41,42] $\tau_1$ and $\tau_2$ reflect electron-hole recombination times via the excitonic emission. They are most probably due to decay of one of the many spectral components that make up the excitonic emission, [1,28] or to a relaxation process within this same manifold. Finally, surely longer time components are present [27,29] but our scans are limited to 100 ps.

The present result of an electron cooling time of ~450 fs is of importance for the description of electron injection times in dye-sensitized ZnO. Indeed, in contrast to dye-sensitized $TiO_2$ where electron injection times are extremely short (< 5 fs) [59,60], in ZnO the injection times are much longer, in the order of several tens of ps. [15,61–63] The present results clearly confirm that they are entirely governed by the dye-ZnO interaction and not by the electron cooling within the ZnO substrate. [63]

Regarding the fs-XANES data, the ~1.4 ps rise of the signal (Figure 6) is to be contrasted with the prompt structural response found in the MD simulations (Figure 7). We conclude from this that the former is mainly determined by the migration of holes, which then localise at singly-charged oxygen vacancies that are expected to be more frequent near the surface of the NP due to a higher density of defects. In order to support this interpretation, we estimated the diffusion time of a hole inside the NP, assuming that it is created at its centre.

Diffusivity (D) is related to charge mobility ($\mu$) via: [64]
$$D = \frac{kT}{q}\mu$$
where k is the Boltzmann constant, T the temperature and q=+e the elementary charge. The diffusion time ($\tau$) and length (L) are related via:
$$L = \sqrt{D\tau}$$
In our case, the largest distance the hole travels in the (32 nm diameter) NP used in the fs-XAS experiment is 16 nm, assuming that the hole is created at the centre of the NP. The values of the hole mobility of ZnO cover a very large range from 0.1 to 50 cm$^2$/V/s, and taking *kT*=25.4 meV at RT, we find that D varies from 0.00254 to 1.27 cm$^2$/s. [2] This implies upper values of migration time of 2 ps to 1 ns. Considering the approximations made in this rough calculation, the fact that a distribution of distances is involved and the over 2-3 orders of magnitude uncertainty in the value of D, this estimate is quite satisfactory. In the context of our hypothesis, it would imply that the above upper value of D is closer to the real value.

The observation that the main features of the 2 ps XANES and EXAFS transients reproduce those of the 100 ps time delay measured at the synchrotron (Figures 3 and 4) can be understood on the basis of figure 7, since the formation of a relaxed cage around the newly formed doubly-charged $O_{vac}$ takes less than 2 ps. In addition, one should stress that there is not only one category of defects and traps and that the 100 ps transient, which was recorded with a 80 ps wide pulse, [30] integrates a much larger sample of structural configurations of the doubly-charged $O_{vac}$'s.

Finally, it is important to stress that the fs-PL and fs-XAS do not monitor the same type of evolution. The ultrafast PL maps the energy relaxation of the electrons in the CB, while the fs-XAS maps the spatial migration of holes in the VB and their trapping. This may include hole energy relaxation but the X-ray observable is not sensitive to it, since it only reports on structural changes at the $O_{vac}$'s that trap the hole. In order to map the energy relaxation of holes, which according to Zhukov et al [41,42] is typically three times faster than the electron relaxation, one would need to detect the holes via ultrafast O K-edge XAS, as was recently reported for TiO$_2$. [37] Since the migration time of the holes is the rate determining step that governs their trapping, it is unlikely that phonon coherences similar to those found in the simulations could be generated. Given the predominance of this optical phonon mode, we should expect it in the steady-state PL spectra. However, as the mode is associated to traps one would expect it in the visible (green) part of the spectrum (Figure S2) since it is associated with hole traps. [30,47,48] However, while the low temperature PL spectrum of ZnO shows rich fine structure of the UV band-gap PL, the green band is featureless. [65] It would be exciting to investigate this phonon mode and its role via impulsive stimulated Raman Spectroscopy.

### III. Conclusions

In summary, we presented a combined ultrafast UV photoluminescence and Zn K-edge absorption study of photoexcited ZnO nanoparticles in solution, complemented by *ab initio* molecular dynamics simulations. Our results shows that electron cooling is ultrafast (<500 fs) and in very good agreement with theoretical predictions. [41,42] The fs X-ray absorption study shows that the signal grows on relatively slow time scales but the transients at the first ps's and at 100 ps are quite similar. In addition, *ab initio* Molecular Dynamics simulations show that upon hole trapping the Zn cage expansion around the doubly-charged oxygen vacancy is prompt and it stabilises within about 2 ps. These results lead us to conclude that the ~1.4 ps rise time of the Zn K-edge signal reflects the diffusion and trapping of holes after they have been created in the regular lattice. This scenario is supported by an estimate of the hole diffusion time in ZnO using literature values of the hole mobility. The processes investigated in this work are summarised in figure 8. The subsequent times found in both the optical PL, the fs-XAS experiments and the deep-UV transient absorption studies are due to electron-hole recombination via both radiative and non-radiative mechanisms, but considering the complex nature of traps in ZnO and their variability with sample preparation, more studies are needed in order to attribute to specific processes.

**Acknowledgments:**

This work was supported by the ERC via the DYNAMOX project, the Swiss SNF via the NCCR:MUST and EPSRC through grant number EP/W008009/1. We acknowledge computational resources from ARCHER2 UK National Computing Service which was granted via HPC-CONEXS, the UK High-End Computing Consortium (EPSRC grant no. EP/X035514/1). The experiment at SACLA was performed with the approval of the Japan Synchrotron Radiation Research Institute (proposal 2014B8039). T. K. acknowledges JSPS KAKENHI (Grant Numbers JP19H05782, JP21H04974, and JP21K18944). J.S. acknowledges the support under the Polish Ministry and Higher Education project nr. 1/SOL/2021/2. NN thanks SNSF for a Marie Heim-Vögtlin grant. G.D., A.M.M, C.S.L. and S.H.S. were supported by the U.S. Department of Energy, Office of Science, Basic Energy Sciences, Chemical Sciences, Geosciences, and Biosciences Division under Contract No. DE-AC02-06CH11357

Table 1: Time constants (all entries are in ps) extracted from the present ultrafast near-UV PL and Zn K-edge XAS experiments and from the previous ps XAS experiment, [30] and ultraviolet transient absorption (TA) spectroscopy. [49] In the PL experiment, $\tau_r$ corresponds to electron cooling in the conduction band, while $\tau_1$ and $\tau_2$ reflect electron-hole recombination via the excitonic emission. In the fs-XAS experiment, $\tau_r$ corresponds to hole migration, trapping time and the cage relaxation at the newly formed doubly-charged vacancy. The longer times are due to electron-hole recombination via the trap PL in the green that is also reported in the ps-XAS experiment. [30] Longer lifetimes have also been reported in the literature. [28,53]

| Measurement | PL (4.66 eV) | Fs UV TA (4.2 eV) [49] | fs-XAS (3.49 eV) | ps-XAS (3.49 eV) [30] |
|---|---|---|---|---|
| $\tau_r$ | 0.42±0.08 | <0.15 | 1.4±0.1 | |
| $\tau_1$ | 6.5± 1 | 1.1 | | |
| $\tau_2$ | 40.0±2 | | | |
| $\tau_3$ | - | 88±1 | 126±44 | 200±130 |
| $\tau_4$ | | 3900±400 | | 1200±300 |

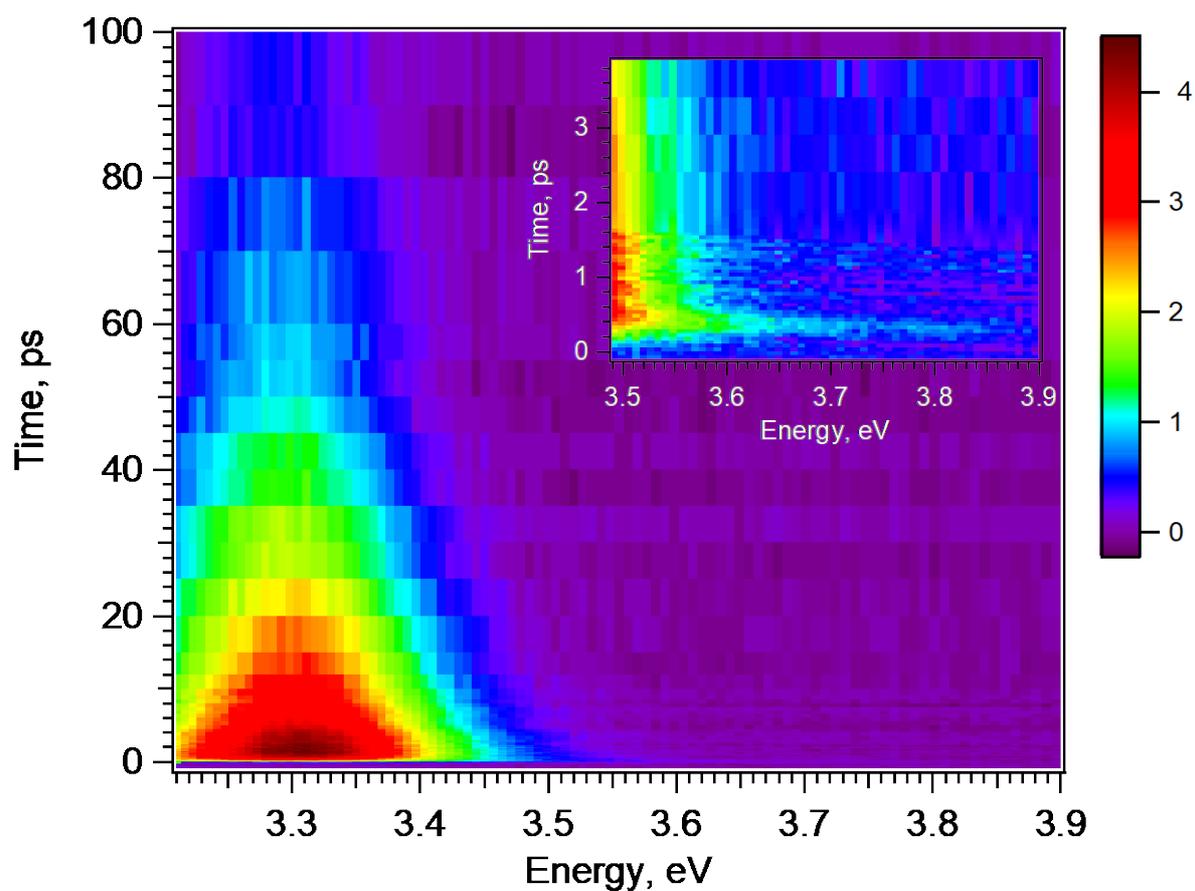

**Figure 1**: Time-energy (t–E) emission plot of the photoluminescence of ZnO nanoparticles in solution, measured upon excitation with 4.66 eV pulses of 5.4 mJ/cm$^2$ fluence. The inset shows the high-energy side of the emission plot over the first 5 ps.

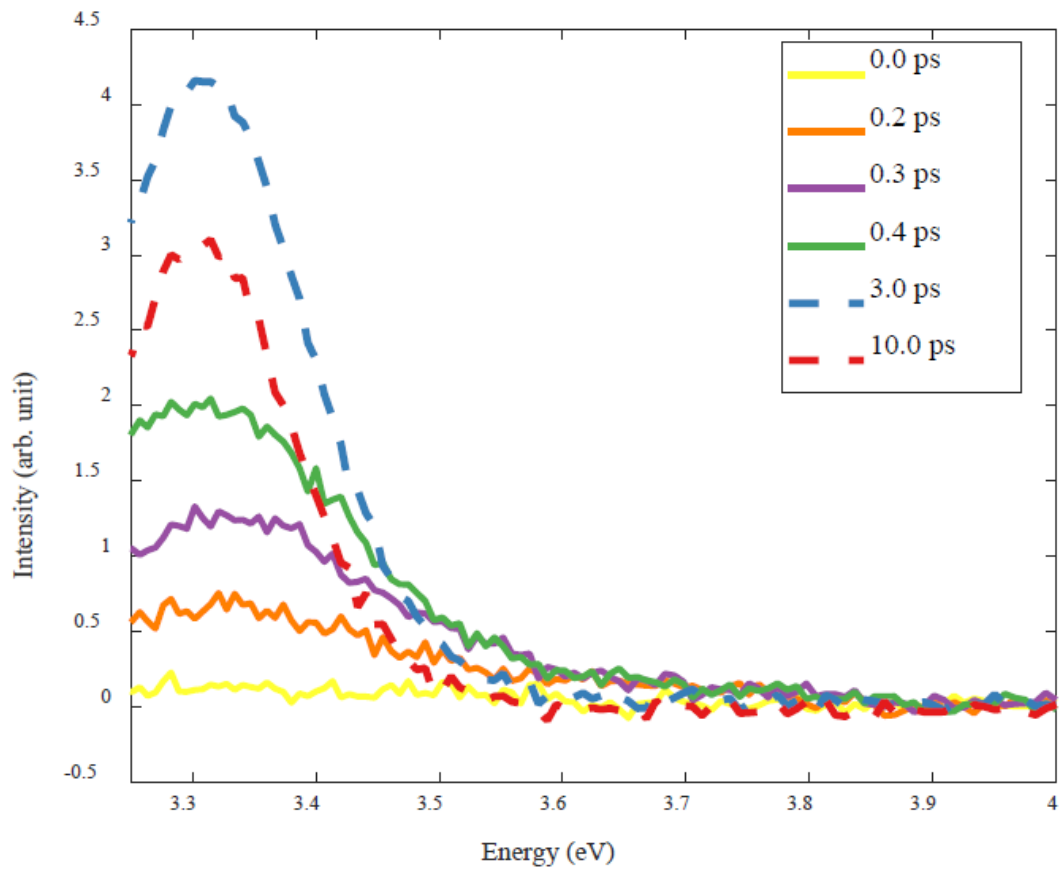

**Figure 2:** Photoluminescence spectra of ZnO nanoparticles in solution, at various time delay after excitation at 4.66 eV and at a fluence of 5.4 mJ/cm$^2$. The spectra show the rise of the band-gap PL band at 3.31 eV. Note the relative contribution of the latter and the shoulder in the 3.45 to 3.7 eV region.

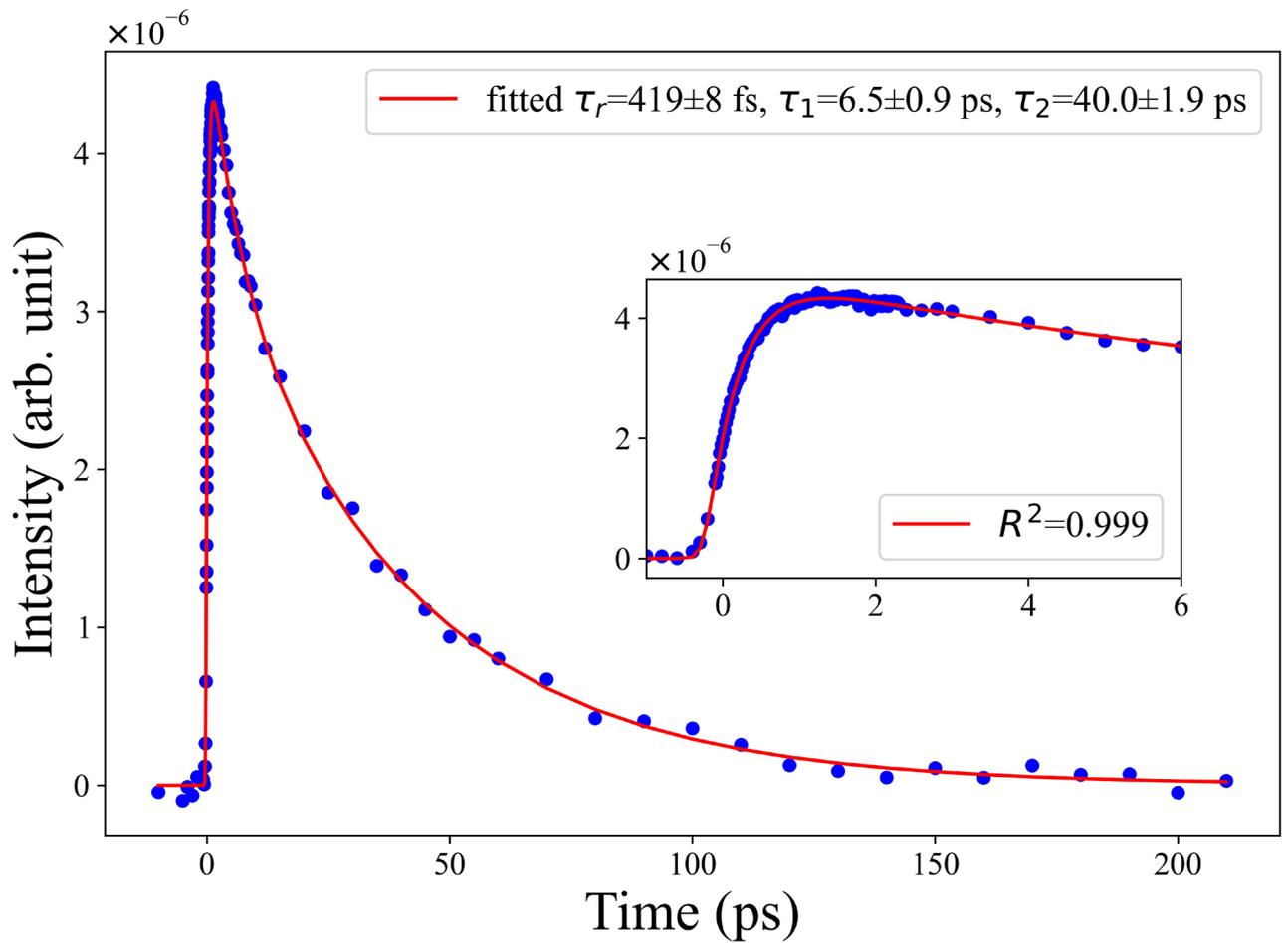

**Figure 3.** Kinetic traces over a time-window of 200 ps of the photoluminescence of ZnO upon 266 nm excitation at a fluence of 5.4 mJ/cm$^2$. The inset shows the time trace over a time window of 6 ps. The fits yield a rise time of approximately 420 fs and decay times of 6.5 ps and 40 ps.

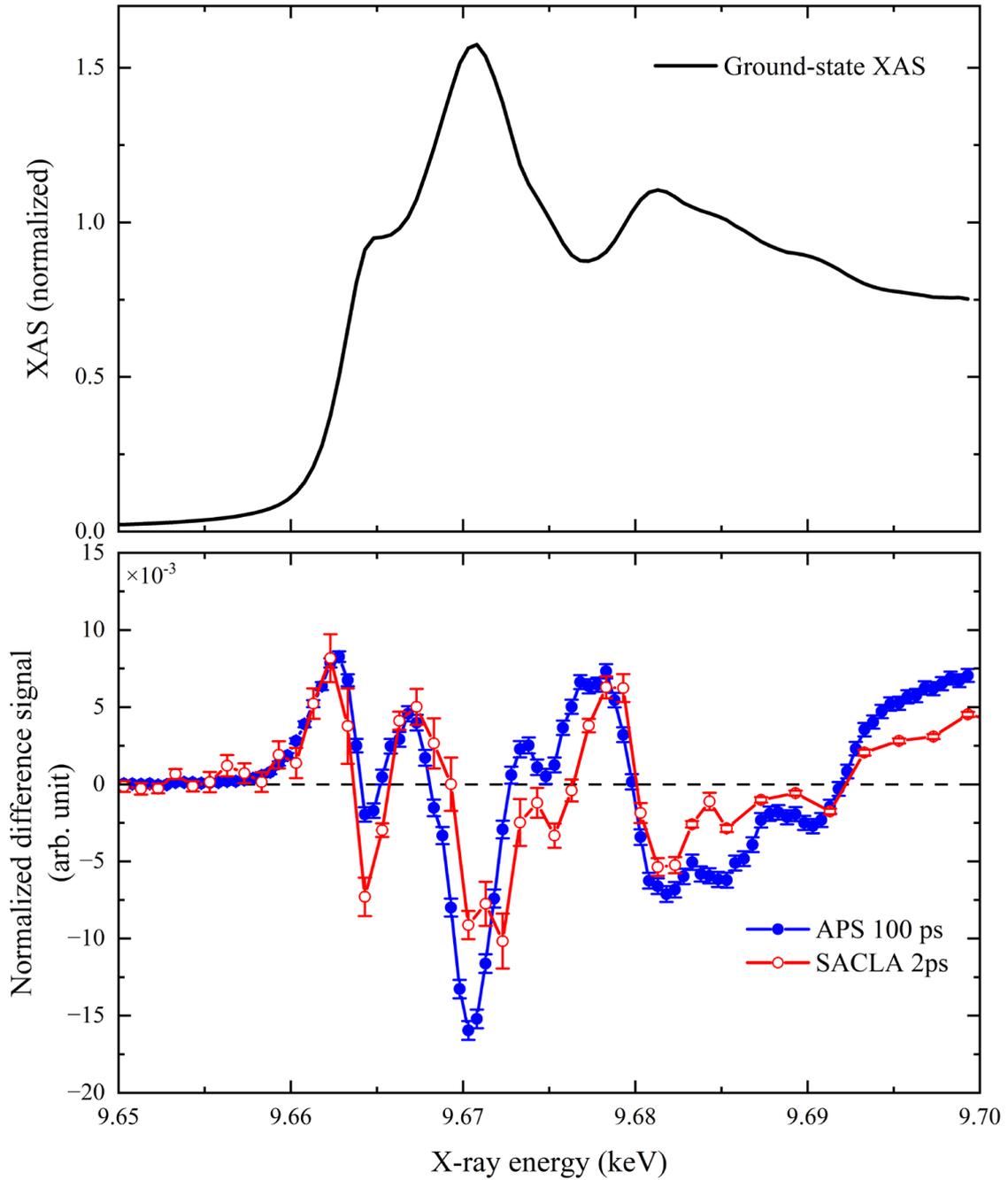

**Figure 4**: Steady-state Zn K-edge spectrum of ZnO nanoparticles in solution (black trace) along with the transient (laser-on minus laser-off) spectrum obtained in ref. [30] using 80 ps resolution (blue points and trace) and the present femtosecond transient (red points and trace). The excitation wavelength for both the ps and the fs transient was 3.49 eV (355 nm).

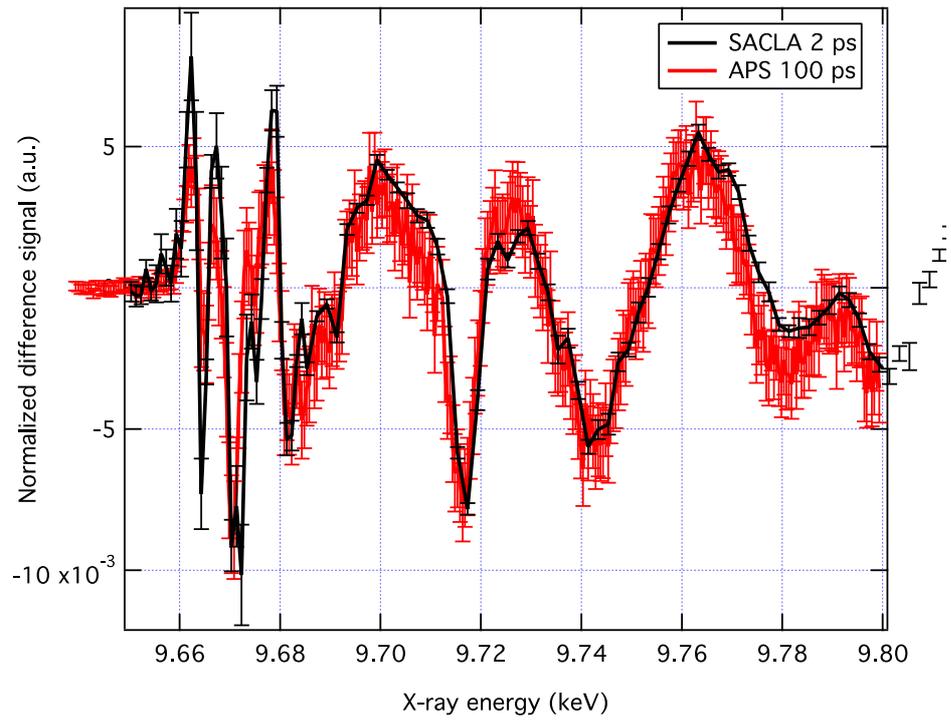

**Figure 5**: Transient X-ray absorption spectrum covering the XANES and EXAFS regions recorded at 100 ps time delay [30] and at 2 ps time delay (this work).

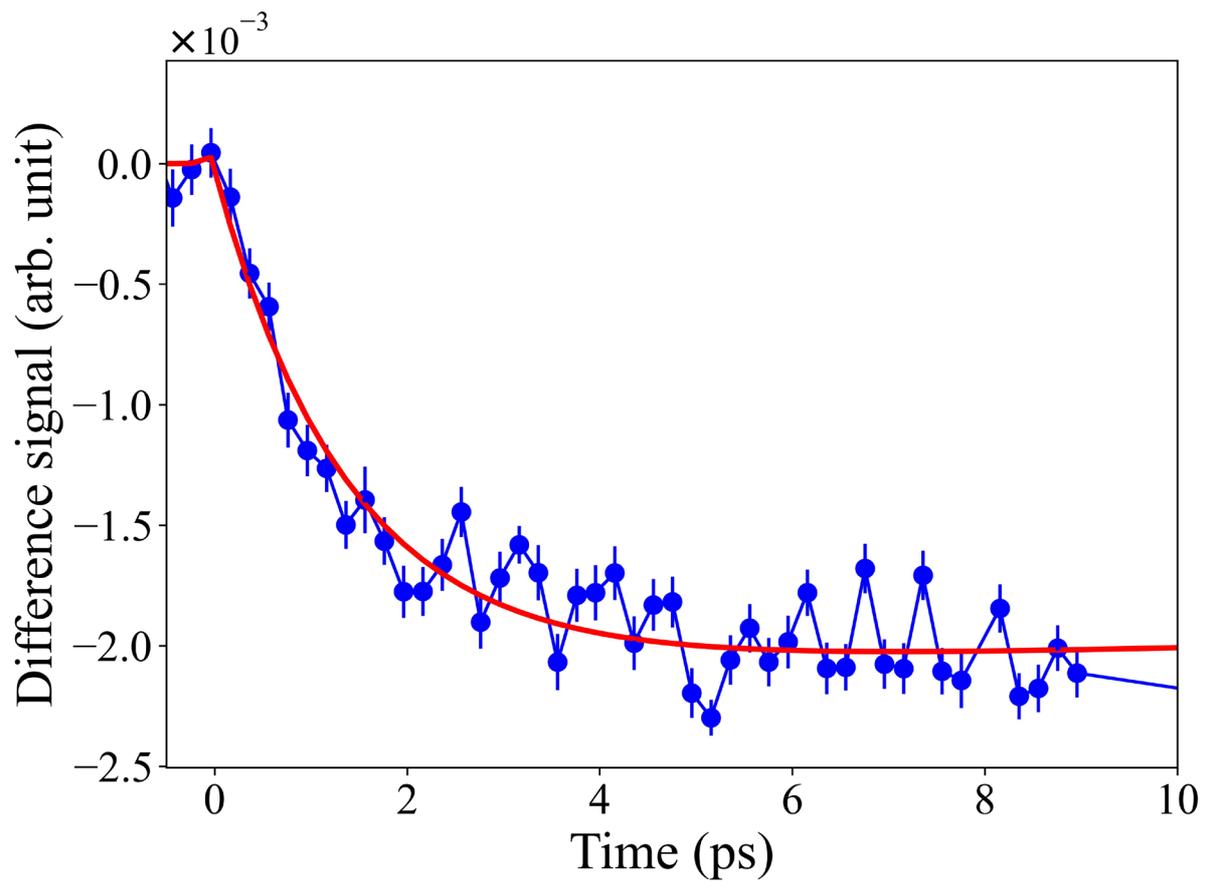

**Figure 6**: Time trace of the signal at 9.67 keV after excitation at 3.49 eV within the first 10 ps. Figure S7 shows the time traces and their fits at intermediate (up to 60 ps) and long (up to 200 ps) time delays.

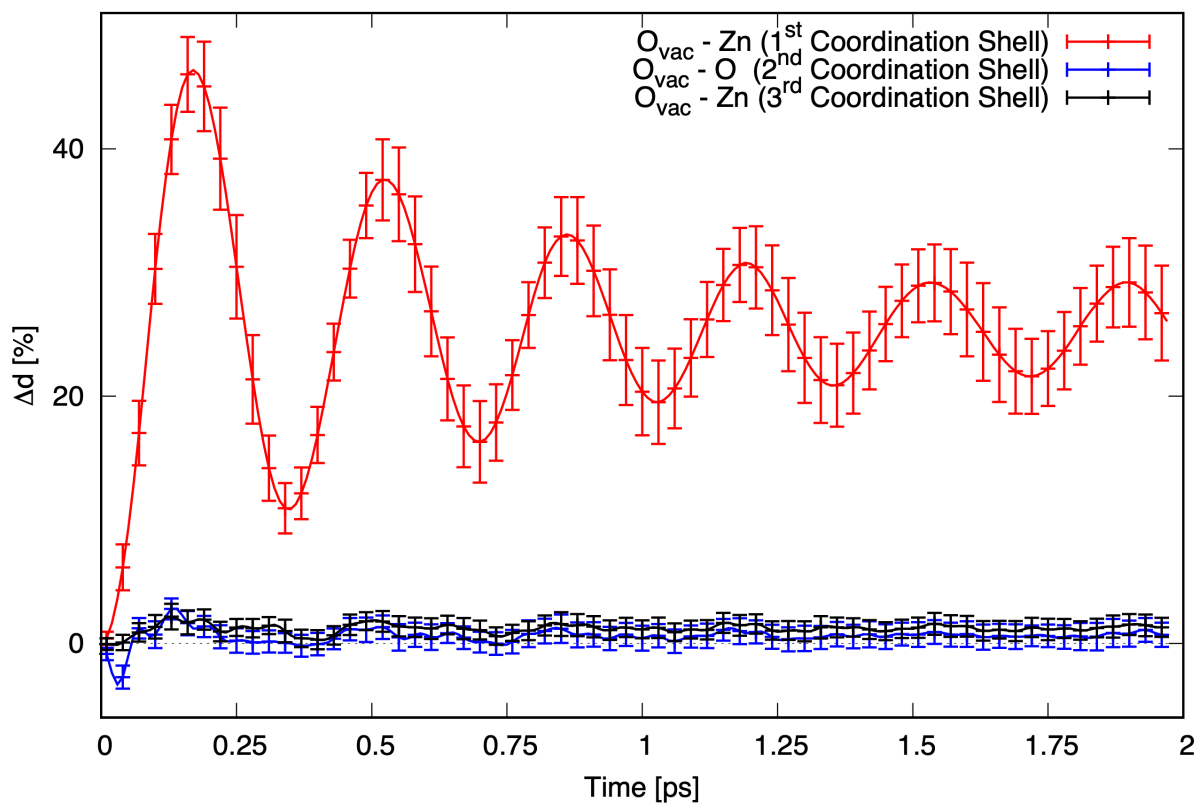

**Figure 7**: Percentage distance change from the oxygen vacancy to the first coordination shell composed of Zn atoms (red), the second coordination shell composed of O atoms (blue) and the third coordination shell composed of Zn atoms (black).

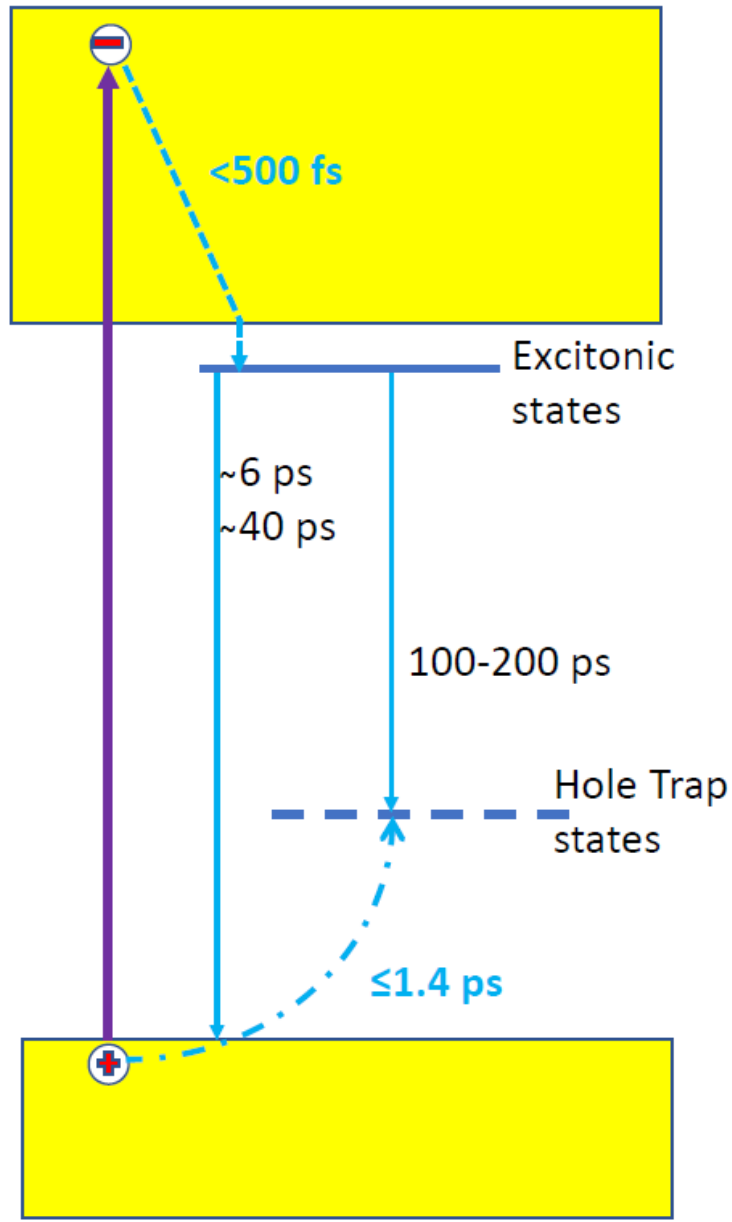

**Figure 8**: Scheme summarising the results and showing the relaxation channels in ZnO for electrons and holes. The dashed line represents energy relaxation, while the dot-dashed line represents diffusion and trapping.

[1] Ü. Özgür, Ya. I. Alivov, C. Liu, A. Teke, M. A. Reshchikov, S. Doğan, V. Avrutin, S.-J. Cho, and H. Morkoç, *A Comprehensive Review of ZnO Materials and Devices*, J. Appl. Phys. **98**, 041301 (2005).

[2] A. Janotti and C. G. Van de Walle, *Fundamentals of Zinc Oxide as a Semiconductor*, Rep. Prog. Phys. **72**, 126501 (2009).

[3] A. Hagfeldt and M. Gratzel, *Molecular Photovoltaics*, Acc. Chem. Res. **33**, 5 (2000).

[4] J. Teuscher, J. C. Brauer, A. Stepanov, A. Solano, A. Boziki, M. Chergui, J.-P. Wolf, U. Rothlisberger, N. Banerji, and J.-E. Moser, *Charge Separation and Carrier Dynamics in Donor-Acceptor Heterojunction Photovoltaic Systems*, Struct. Dyn. **4**, 061503 (2017).

[5] A. L. Linsebigler, G. Q. Lu, and J. T. Yates, *Photocatalysis on Tio2 Surfaces - Principles, Mechanisms, and Selected Results*, Chem. Rev. **95**, 3 (1995).

[6] K. Nakata and A. Fujishima, *TiO2 Photocatalysis: Design and Applications*, J. Photochem. Photobiol. C Photochem. Rev. **13**, 3 (2012).

[7] K. M. Lee, C. W. Lai, K. S. Ngai, and J. C. Juan, *Recent Developments of Zinc Oxide Based Photocatalyst in Water Treatment Technology: A Review*, Water Res. **88**, 428 (2016).

[8] J. Ji, A. Colosimo, W. Anwand, L. Boatner, A. Wagner, P. Stepanov, T. Trinh, M. Liedke, R. Krause-Rehberg, and T. Cowan, *ZnO Luminescence and Scintillation Studied via Photoexcitation, X-Ray Excitation, and Gamma-Induced Positron Spectroscopy*, Sci. Rep. **6**, 31238 (2016).

[9] H. Hosono and K. Ueda, *Transparent Conductive Oxides*, in *Springer Handbook of Electronic and Photonic Materials*, edited by S. Kasap and P. Capper (Springer International Publishing, Cham, 2017), pp. 1–1.

[10] H. Dong, B. Zhou, J. Li, J. Zhan, and L. Zhang, *Ultraviolet Lasing Behavior in ZnO Optical Microcavities*, J. Materiomics **3**, 4 (2017).

[11] E. Baldini, T. Palmieri, A. Dominguez, P. Ruello, A. Rubio, and M. Chergui, *Phonon-Driven Selective Modulation of Exciton Oscillator Strengths in Anatase TiO2 Nanoparticles*, Nano Lett. **18**, 8 (2018).

[12] E. Baldini, A. Dominguez, T. Palmieri, O. Cannelli, A. Rubio, P. Ruello, and M. Chergui, *Exciton Control in a Room Temperature Bulk Semiconductor with Coherent Strain Pulses*, Sci. Adv. **5**, 11 (2019).

[13] J. J. Cavaleri, D. E. Skinner, D. P. Colombo, and R. M. Bowman, *Femtosecond Study of the Size-Dependent Charge-Carrier Dynamics in Zno Nanocluster Solutions*, J. Chem. Phys. **103**, 13 (1995).

[14] D. P. Colombo, K. A. Roussel, J. Saeh, D. E. Skinner, J. J. Cavaleri, and R. M. Bowman, *Femtosecond Study of the Intensity Dependence of Electron-Hole Dynamics in Tio2 Nanoclusters*, Chem. Phys. Lett. **232**, 3 (1995).

[15] H. Nemec, P. Kuzel, and V. Sundstrom, *Far-Infrared Response of Free Charge Carriers Localized in Semiconductor Nanoparticles*, Phys. Rev. B **79**, 11 (2009).

[16] H. Nemec, P. Kuzel, and V. Sundstrom, *Charge Transport in Nanostructured Materials for Solar Energy Conversion Studied by Time-Resolved Terahertz Spectroscopy*, J. Photochem. Photobiol. -Chem. **215**, 2–3 (2010).

[17] C. Bauer, G. Boschloo, E. Mukhtar, and A. Hagfeldt, *Ultrafast Relaxation Dynamics of Charge Carriers Relaxation in ZnO Nanocrystalline Thin Films*, Chem. Phys. Lett. **387**, 1–3 (2004).

[18] E. Hendry, F. Wang, J. Shan, T. F. Heinz, and M. Bonn, *Electron Transport in TiO2 Probed by THz Time-Domain Spectroscopy*, Phys. Rev. B **69**, 8 (2004).

[19] A. Furube, Y. Tarnaki, M. Murai, K. Hara, R. Katoh, and M. Tachiya, *Femtosecond Visible-to-IR Spectroscopy of TiO2 Nanocrystalline Films: Dynamics of UV-Generated*

*Charge Carrier Relaxation at Different Excitation Wavelengths - Art. No. 66430J*, Phys. Chem. Interfaces Nanomater. Vi **6643**, J6430 (2007).

[20] A. Schleife, C. Rödl, F. Fuchs, K. Hannewald, and F. Bechstedt, *Optical Absorption in Degenerately Doped Semiconductors: Mott Transition or Mahan Excitons?*, Phys. Rev. Lett. **107**, 23 (2011).

[21] S. Richter et al., *Ultrafast Dynamics of Hot Charge Carriers in an Oxide Semiconductor Probed by Femtosecond Spectroscopic Ellipsometry*, New J. Phys. **22**, 8 (2020).

[22] O. Herrfurth et al., *Transient Birefringence and Dichroism in ZnO Studied with Fs-Time-Resolved Spectroscopic Ellipsometry*, Phys. Rev. Res. **3**, 013246 (2021).

[23] A. Yamamoto, T. Kido, T. Goto, Y. Chen, T. Yao, and A. Kasuya, *Dynamics of Photoexcited Carriers in ZnO Epitaxial Thin Films*, Appl. Phys. Lett. **75**, 469 (1999).

[24] C.-K. Sun, S.-Z. Sun, K.-H. Lin, K. Y.-J. Zhang, H.-L. Liu, S.-C. Liu, and J.-J. Wu, *Ultrafast Carrier Dynamics in ZnO Nanorods*, Appl. Phys. Lett. **87**, 023106 (2005).

[25] X. M. Wen et al., *Ultrafast Dynamics in ZnO/ZnMgO Multiple Quantum Wells*, Nanotechnology **18**, 315403 (2007).

[26] E. Baldini et al., *Strongly Bound Excitons in Anatase $TiO_2$ Single Crystals and Nanoparticles*, Nat. Commun. **8**, 1 (2017).

[27] E. Baldini, T. Palmieri, E. Pomarico, G. Auböck, and M. Chergui, *Clocking the Ultrafast Electron Cooling in Anatase Titanium Dioxide Nanoparticles*, ACS Photonics **5**, 4 (2018).

[28] L. Foglia, S. Vempati, B. Tanda Bonkano, L. Gierster, M. Wolf, S. Sadofev, and J. Stähler, *Revealing the Competing Contributions of Charge Carriers, Excitons, and Defects to the Non-Equilibrium Optical Properties of ZnO*, Struct. Dyn. **6**, 034501 (2019).

[29] C. Bonati, M. B. Mohamed, D. Tonti, G. Zgrablic, S. Haacke, F. van Mourik, and M. Chergui, *Spectral and Dynamical Characterization of Multiexcitons in Colloidal CdSe Semiconductor Quantum Dots*, Phys. Rev. B **71**, 20 (2005).

[30] T. J. Penfold et al., *Revealing Hole Trapping in Zinc Oxide Nanoparticles by Time-Resolved X-Ray Spectroscopy*, Nat. Commun. **9**, 1 (2018).

[31] M. H. Rittmann-Frank, C. J. Milne, J. Rittmann, M. Reinhard, T. J. Penfold, and M. Chergui, *Mapping of the Photoinduced Electron Traps in TiO2 by Picosecond X-Ray Absorption Spectroscopy*, Angew. Chem. Int. Ed. **53**, 23 (2014).

[32] F. G. Santomauro et al., *Localized Holes and Delocalized Electrons in Photoexcited Inorganic Perovskites: Watching Each Atomic Actor by Picosecond X-Ray Absorption Spectroscopy*, Struct. Dyn. **4**, 044002 (2017).

[33] Y. Obara et al., *Femtosecond Time-Resolved X-Ray Absorption Spectroscopy of Anatase TiO2 Nanoparticles Using XFEL*, Struct. Dyn. **4**, 044033 (2017).

[34] J. Budarz, F. G. Santomauro, M. H. Rittmann-Frank, C. J. Milne, T. Huthwelker, D. Grolimund, J. Rittmann, D. Kinschel, T. Rossi, and M. Chergui, *Time-Resolved Element-Selective Probing of Charge Carriers in Solar Materials*, Chimia **71**, 11 (2017).

[35] Y. Uemura, T. Yokoyama, T. Katayama, S. Nozawa, and K. Asakura, *Tracking the Local Structure Change during the Photoabsorption Processes of Photocatalysts by the Ultrafast Pump-Probe XAFS Method*, Appl. Sci. **10**, 21 (2020).

[36] A. S. M. Ismail et al., *Direct Observation of the Electronic States of Photoexcited Hematite with Ultrafast 2p3d X-Ray Absorption Spectroscopy and Resonant Inelastic X-Ray Scattering*, Phys. Chem. Chem. Phys. **22**, 5 (2020).

[37] S. H. Park et al., *Direct and Real-Time Observation of Hole Transport Dynamics in Anatase TiO2 Using X-Ray Free-Electron Laser*, Nat. Commun. **13**, 1 (2022).

[38] R. Asahi, Y. Taga, W. Mannstadt, and A. J. Freeman, *Electronic and Optical Properties of Anatase TiO2*, Phys. Rev. B **61**, 11 (2000).

[39] D. C. Reynolds, D. C. Look, B. Jogai, C. W. Litton, G. Cantwell, and W. C. Harsch, *Valence-Band Ordering in ZnO*, Phys. Rev. B **60**, 2340 (1999).

[40] W. R. Lambrecht, A. V. Rodina, S. Limpijumnong, B. Segall, and B. K. Meyer, *Valence-Band Ordering and Magneto-Optic Exciton Fine Structure in ZnO*, Phys. Rev. B **65**, 7 (2002).

[41] V. P. Zhukov, P. M. Echenique, and E. V. Chulkov, *Two Types of Excited Electron Dynamics in Zinc Oxide*, Phys. Rev. B **82**, 094302 (2010).

[42] V. P. Zhukov, V. G. Tyuterev, E. V. Chulkov, and P. M. Echenique, *Hole-Phonon Relaxation and Photocatalytic Properties of Titanium Dioxide and Zinc Oxide: First-Principles Approach*, Int. J. Photoenergy **2014**, e738921 (2014).

[43] T. Rossi, T. J. Penfold, M. H. Rittmann-Frank, M. Reinhard, J. Rittmann, C. N. Borca, D. Grolimund, C. J. Milne, and M. Chergui, *Characterizing the Structure and Defect Concentration of ZnO Nanoparticles in a Colloidal Solution*, J. Phys. Chem. C **118**, 19422 (2014).

[44] A. Janotti and C. G. Van de Walle, *Oxygen Vacancies in ZnO*, Appl. Phys. Lett. **87**, 12 (2005).

[45] A. Janotti and C. G. Van de Walle, *Native Point Defects in ZnO*, Phys. Rev. B **76**, 16 (2007).

[46] A. Cannizzo, O. Bram, G. Zgrablic, A. Tortschanoff, A. A. Oskouei, F. van Mourik, and M. Chergui, *Femtosecond Fluorescence Upconversion Setup with Broadband Detection in the Ultraviolet*, Opt. Lett. **32**, 24 (2007).

[47] K. Vanheusden, W. L. Warren, C. H. Seager, D. R. Tallant, J. A. Voigt, and B. E. Gnade, *Mechanisms behind Green Photoluminescence in ZnO Phosphor Powders*, J. Appl. Phys. **79**, 7983 (1996).

[48] K. Vanheusden, C. H. Seager, W. L. Warren, D. R. Tallant, and J. A. Voigt, *Correlation between Photoluminescence and Oxygen Vacancies in ZnO Phosphors*, Appl. Phys. Lett. **68**, 403 (1996).

[49] E. Baldini, T. Palmieri, T. Rossi, M. Oppermann, E. Pomarico, G. Auböck, and M. Chergui, *Interfacial Electron Injection Probed by a Substrate-Specific Excitonic Signature*, J. Am. Chem. Soc. **139**, 33 (2017).

[50] W. Gawelda, V.-T. Pham, R. M. van der Veen, D. Grolimund, R. Abela, M. Chergui, and C. Bressler, *Structural Analysis of Ultrafast Extended X-Ray Absorption Fine Structure with Subpicometer Spatial Resolution: Application to Spin Crossover Complexes*, J. Chem. Phys. **130**, 124520 (2009).

[51] M. Chergui, *Picosecond and Femtosecond X-Ray Absorption Spectroscopy of Molecular Systems*, Acta Crystallogr. Sect. A **66**, 229 (2010).

[52] L. X. Chen, *X-Ray Transient Absorption Spectroscopy*, in *X-Ray Absorption and X-Ray Emission Spectroscopy* (John Wiley & Sons, Ltd, 2016), pp. 213–249.

[53] T. C. Rossi, C. P. Dykstra, T. N. Haddock, R. Wallick, J. H. Burke, C. M. Gentle, G. Doumy, A. M. March, and R. M. van der Veen, *Charge Carrier Screening in Photoexcited Epitaxial Semiconductor Nanorods Revealed by Transient X-Ray Absorption Linear Dichroism*, Nano Lett. **21**, 9534 (2021).

[54] R. C. Rai, *Analysis of the Urbach Tails in Absorption Spectra of Undoped ZnO Thin Films*, J. Appl. Phys. **113**, 153508 (2013).

[55] R. Cuscó, E. Alarcón-Lladó, J. Ibáñez, L. Artús, J. Jiménez, B. Wang, and M. J. Callahan, *Temperature Dependence of Raman Scattering in ZnO*, Phys. Rev. B **75**, 165202 (2007).

[56] B. K. Meyer et al., *Bound Exciton and Donor–Acceptor Pair Recombinations in ZnO*, Phys. Status Solidi B **241**, 231 (2004).

[57] J. C. Johnson, K. P. Knutsen, H. Yan, M. Law, Y. Zhang, P. Yang, and R. J. Saykally, *Ultrafast Carrier Dynamics in Single ZnO Nanowire and Nanoribbon Lasers*, Nano Lett. **4**, 197 (2004).

[58] J. C. Deinert, D. Wegkamp, M. Meyer, C. Richter, M. Wolf, and J. Stahler, *Ultrafast Exciton Formation at the ZnO(10(1)over-Bar0) Surface*, Phys. Rev. Lett. **113**, 5 (2014).

[59] R. Huber, J. E. Moser, M. Gratzel, and J. Wachtveitl, *Real-Time Observation of Photoinduced Adiabatic Electron Transfer in Strongly Coupled Dye/Semiconductor Colloidal Systems with a 6 Fs Time Constant*, J. Phys. Chem. B **106**, 25 (2002).

[60] O. Bram, A. Cannizzo, and M. Chergui, *Ultrafast Fluorescence Studies of Dye Sensitized Solar Cells*, Phys. Chem. Chem. Phys. **14**, 22 (2012).

[61] G. Benko, J. Kallioinen, J. E. I. Korppi-Tommola, A. P. Yartsev, and V. Sundstrom, *Photoinduced Ultrafast Dye-to-Semiconductor Electron Injection from Nonthermalized and Thermalized Donor States*, J. Am. Chem. Soc. **124**, 3 (2002).

[62] P. Myllyperkio, G. Benko, J. Korppi-Tommola, A. P. Yartsev, and V. Sundstrom, *A Study of Electron Transfer in Ru(Dcbpy)(2)(NCS)(2) Sensitized Nanocrystalline TiO2 and SnO2 Films Induced by Red-Wing Excitation*, Phys. Chem. Chem. Phys. **10**, 7 (2008).

[63] H. Nemec, J. Rochford, O. Taratula, E. Galoppini, P. Kuzel, T. Polivka, A. Yartsev, and V. Sundstrom, *Influence of the Electron-Cation Interaction on Electron Mobility in Dye-Sensitized ZnO and TiO2 Nanocrystals: A Study Using Ultrafast Terahertz Spectroscopy*, Phys. Rev. Lett. **104**, 19 (2010).

[64] E. Flitsyian, Z. Dashevsky, and L. Chernyak, *Minority Carrier Transport in ZnO and Related Materials*, in *GaN and ZnO-Based Materials and Devices*, edited by S. Pearton (Springer, Berlin, Heidelberg, 2012), pp. 317–347.

[65] B. K. Meyer et al., *Bound Exciton and Donor–Acceptor Pair Recombinations in ZnO*, Phys. Status Solidi B **241**, 231 (2004).

# Supplementary Material section

## Disentangling the Evolution of Electrons and Holes in photoexcited ZnO Nanoparticles

Christopher J. Milne[1,2]*, Natalia Nagornova[3], Thomas Pope[4], Hui-Yuan Chen[3], Thomas Rossi[3], Jakub Szlachetko[2,5], Wojciech Gawelda[1,5,6], Alexander Britz[1,7], Tim B. van Driel[8], Leonardo Sala[2], Simon Ebner[2], Tetsuo Katayama[9,10], Stephen H. Southworth[11], Gilles Doumy[11], Anne Marie March[11], C. Stefan Lehmann[11,12], Melanie Mucke[13], Denys Iablonskyi[14], Yoshiaki Kumagai[14], Gregor Knopp[2], Koji Motomura[14], Tadashi Togashi[9], Shigeki Owada[10], Makina Yabashi[10], Martin M. Nielsen[8], Marek Pajek[5], Kiyoshi Ueda[10,14], Rafael Abela[2], Thomas J. Penfold[4] and Majed Chergui[3]*

1 European XFEL, D-22761 Hamburg, Germany
2 SwissFEL, Paul-Scherrer-Institute, 5232 Villigen-PSI, Switzerland
3 Lausanne Centre for Ultrafast Science (LACUS), ISIC, FSB, Ecole Polytechnique Fédérale de Lausanne, CH-1015 Lausanne, Switzerland
4 Chemistry – School of Natural and Environmental Sciences, Newcastle University, Newcastle upon Tyne, NE1 7RU, UK
5 SOLARIS National Synchrotron Radiation Centre, Jagellonian University, 30-387 Krakow, Poland
6 IMDEA Nanoscience Institute, Calle Faraday 9, Campus Cantoblanco, 28049 Madrid, SPAIN
7 The Hamburg Centre for Ultrafast Imaging, Luruper Chaussee 149, 22761 Hamburg, Germany
8 Department of Physics, Technical University of Denmark, 2800 Kongens Lyngby, Denmark
9 Japan Synchrotron Radiation Research Institute (JASRI), Kouto 1-1-1, Sayo, Hyogo 679-5198, Japan
10 RIKEN SPring-8 Center, Kouto 1-1-1, Sayo, Hyogo 679-5148, Japan
11 Argonne National Laboratory, 9700 S. Cass Ave., Argonne, Illinois 60439, USA
12 Advanced Research Center for Nanolithography (ARCNL), Science Park 106, 1098 XG Amsterdam, Netherlands
13 Department of Physics and Astronomy, Uppsala University, 751 20 Uppsala, Sweden
14 Institute of Multidisciplinary Research for Advanced Materials, Tohoku University, Sendai 980-8577, Japan

*Corresponding authors:* christopher.milne@xfel.eu; majed.chergui@epfl.ch

### S.1. Experimental set-up and procedures:

*S.1.1 Samples:*

For the photoluminescence up-conversion studies, 30-50 nm diameter ZnO nanoparticles from Sigma Aldrich (40 wt. % in ethanol) and were further dispersed in ethanol (Fischer Scientific, analytical reagent grade). The concentration of ZnO NPs was adapted to achieve an OD of 0.3 at 266 nm in a 200 μm-thick flow cell. In order to minimize damage, the sample was flowed and the sample cell translated horizontally, to ensure a fresh spot for excitation.

For the X-ray experiments, the sample was a 170 mM dispersion of 32 nm diameter ZnO nanoparticles in water. It was prepared from a commercially available sample (Sigma-Aldrich 721077) which we have previously characterized in separate publications. [1,2] Under the conditions of the experiment, the optical transmission of the jet was measured to be 33%, corresponding to an absorbance of 0.5.

*S.1.2 Fluorescence up-conversion set-up*

Time-resolved photoluminescence (PL) spectra of ZnO nanoparticles (NPs) were recorded using a broadband fluorescence up-conversion set-up, described in detail in ref. [3] Briefly, 800 nm pulses of 4 µJ typical energy are provided by a Coherent-RegA Ti:Sapphire regenerative amplifier operated at 150 kHz repetition rate. One third of the 800 nm beam is used to obtain 4.66 eV excitation pulses through a third harmonic generation process, while the other two thirds are used as a gate pulse, which is sent to the up-conversion β-barium borate (BBO) crystal through an optical delay stage. The 4.66 eV excitation beam of about 20 nJ per pulse is focused down to a 20-µm spot inside of a 0.2-mm thick UV-grade fused silica flow cell (Starna), where the sample is continuously circulated using a peristaltic pump. Emission from the sample is collected by a wide-angle parabolic mirror and directed to a second mirror that focuses it onto the 250-µm thick BBO crystal, where it is mixed with the delayed gate beam in a slightly non-collinear geometry to produce an up-converted signal. The up-converted signal is then spatially filtered and detected with a spectrograph and a CCD (charge-coupled device) camera. To accomplish a broadband detection, the up-conversion crystal is rotated with a constant angular speed during the integration time of the CCD camera in order to phase match a wide spectral region at each fixed time delay. The instrument response function (IRF) of the set-up is well represented by a Gaussian with FWHM~230 fs from the Raman line of $H_2O$ excited at 266 nm (Figure S6). The kinetic trace of ZnO PL recorded using a single-wavelength detection at 3.31 eV is also shown in Figure S6. We verified that the fluence dependence of the PL signal upon 4.66 eV excitation for the range of fluences used here is linear (Figures S2 to S4). Figure S7 shows the time trace of the PL over a longer time window.

*S.1.3 Femtosecond X-ray absorption spectroscopy*

The experiments were performed at beamline 3 (BL3) of SACLA. [4–7] SACLA was operated with mean pulse energy of 240 µJ, pulse duration of 20 fs, and operation frequency of 10 Hz during the experiment. Concerning the timing of the laser system, a synchronization system was used with direct detection of an optical pulse train, yielding a time jitter of approximately 300 fs.

The sample was excited using a 355 nm (3.49 eV) pulse having a temporal width of ~40 fs and an energy/pulse of 21 μJ. The laser pulses were focussed to 40 μm FWHM on the sample. Measurements at lower fluence measurements were also carried out showing no variation and the results reported here concern the above numbers.

## S.2. Fit of kinetic traces:

The fit of the kinetic traces was conducted using multiexponential functions composed of a rising component and decay ones, convoluted to the IRF of the experiments, which is 230 fs for the fs optical photoluminescence (fs-PL), and 150 fs for the fs-XANES measurements.

The analytic formula for the rise component is expressed as follow:

$$\left[a_r\left(1 - e^{-\frac{t}{\tau_r}}\right)\right] \otimes IRF(t) = \frac{1}{2}a_r\left[1 + erf\left(\frac{t}{\sqrt{2}\sigma}\right)\right] - \frac{1}{2}a_r exp\left[\frac{(\mu - t)}{\tau_r} + \frac{\sigma^2}{2\tau_r^2}\right] \times \left[1 + erf\left(\frac{t - \left(\mu + \frac{\sigma^2}{\tau_r}\right)}{\sqrt{2}\sigma}\right)\right]$$

while that of the decay term is:

$$a_i\left(e^{-\frac{t}{\tau_i}}\right) \otimes IRF(t) = \frac{1}{2}a_i exp\left[\frac{(\mu - t)}{\tau_i} + \frac{\sigma^2}{2\tau_i^2}\right] \times \left[1 + erf\left(\frac{t - \left(\mu + \frac{\sigma^2}{\tau_i}\right)}{\sqrt{2}\sigma}\right)\right]$$

$a_r$ and $a_i$ are the pre-exponential factors for the rise and decay terms, respectively. $\tau_r$ is the risetime constant; $\tau_i$ is the decay time constant; $\sigma$ represents the width of IRF, where $IRF(FWHM) = 2\sqrt{\ln(2)}\,\sigma$.

For fs-PL, one risetime and two decay components were used (figure S7). For the fs-XANES, we considered two cases for the fits: (a) one rise and one decay component and; (b) one rise and two decay components. The fits are shown in figures S8 and S9, respectively. They are both of similar and the fit parameters are listed in table S2.

## S.3. Ab-initio Molecular Dynamics simulations:

All density functional theory (DFT) calculations were performed with the CASTEP Simulation package, [8] with an energy cutoff of 400 eV, 3 k-points along each supercell vector, and using the PBE exchange-correlation functional. [9] The supercell consisted of an optimized ZnO crystal structure with four unit cells in all directions. Once optimized, *ab initio* molecular dynamics (AIMD) was performed for 2 ps at 300 K. 128 snapshots were taken and an oxygen atom was removed to create a group of starting geometries containing an oxygen vacancy. Additional AIMD calculations were performed on the structures with an oxygen vacancy and a 2+ charge. We calculated the average distance from the vacancy and the nearest 4 Zn atoms, nearest 12 oxygen atoms, and next-nearest 12 Zn atoms (corresponding to the 1st, 2nd and 3rd coordination shells, respectively), for each time step. Finally, the average displacement for each coordination shell was taken over all systems for each time step.

**Table S1**. Time constants of the emission of ZnO NPs excited with 266 nm, obtained using multiexponential fits of the kinetic traces, extracted from the (t-E) plots at max of the emission (shown in Fig. 1). Rise time components ($\tau_r$) for different fluence values were determined using multiexponential fits, while $\tau_1$ and $\tau_2$ were kept fixed to the values determined by fitting the kinetic trace measured at the max fluence (5.4 mJ/cm$^2$). This is the only one for which a long-time scan was recorded.

| Fluence (mJ/cm$^2$) | $\tau_r$ (fs) | $\tau_1$ (ps) | $\tau_2$ (ps) |
|---|---|---|---|
| 0.8 | 340 ± 150 fs | - | - |
| 2.3 | 494 ± 81 fs | - | - |
| 4.7 | 451 ± 43 fs | - | - |
| 5.4 | 419± 8 fs | 6.5 ±0.9 ps | 40.0 ± 1.9 ps |

**Table S2**: Summary of fit parameter of the XANES time traces (Figures S7 and S8). The fit formulae are presented in § S.2. The preexponential factors are given in the bottom three lines (not normalized).

|  | fs-PL | fs-XANES (1rising+1decay) |
|---|---|---|
| $\tau_r$ (ps) | 419±8 (fs) | 1.4 ± 0.1 |
| $\tau_1$ (ps) | 6.5±0.9 | 126.5 ± 43.7 |
| $\tau_2$ (ps) | 40.0±1.9 | --- |
| $a_r$ | 4.87 x 10$^{-6}$ | -2.28 x 10$^{-3}$ |
| $a_1$ | 1.33 x 10$^{-6}$ | -1.25 x 10$^{-3}$ |
| $a_2$ | 3.53 x 10$^{-6}$ | --- |

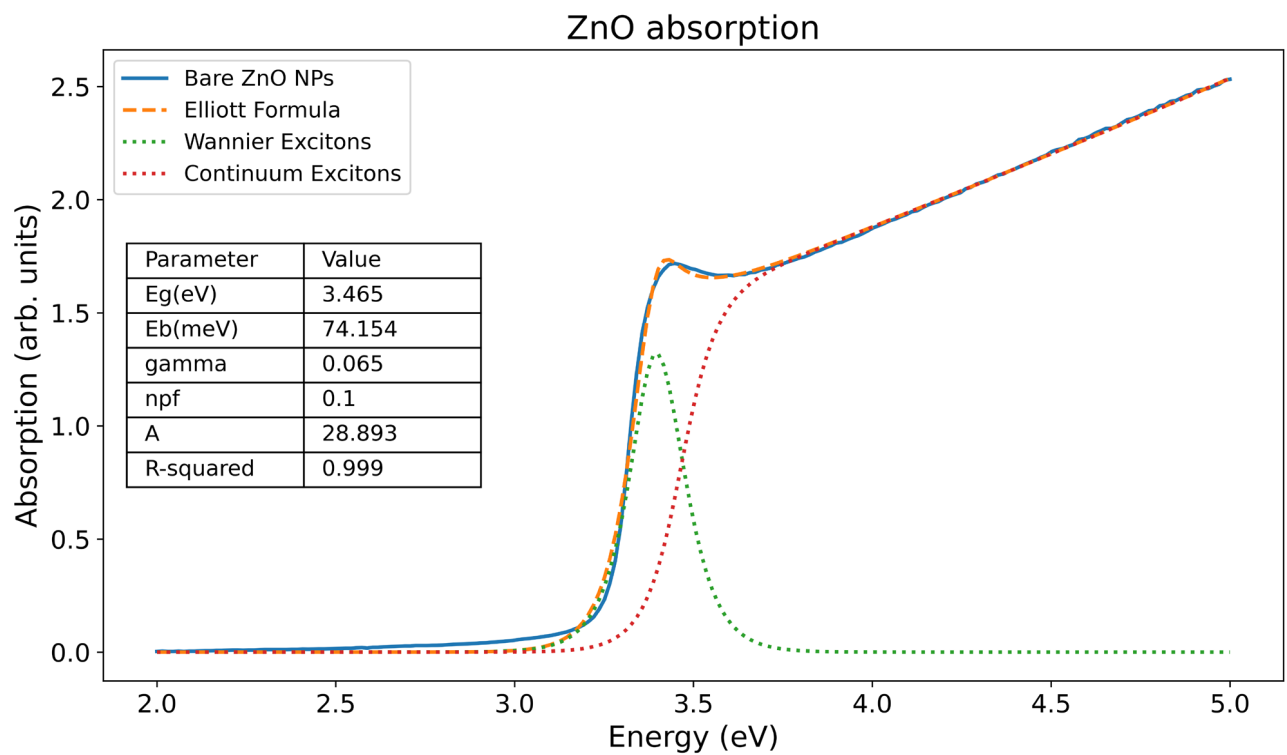

Figure S1: Steady-state absorption of our sample of ZnO NPs in solution and fit of the absorption threshold using the Elliott formula and a Wannier exciton band, assuming a binding energy of 74 meV and an electronic gap of 3.465 eV.

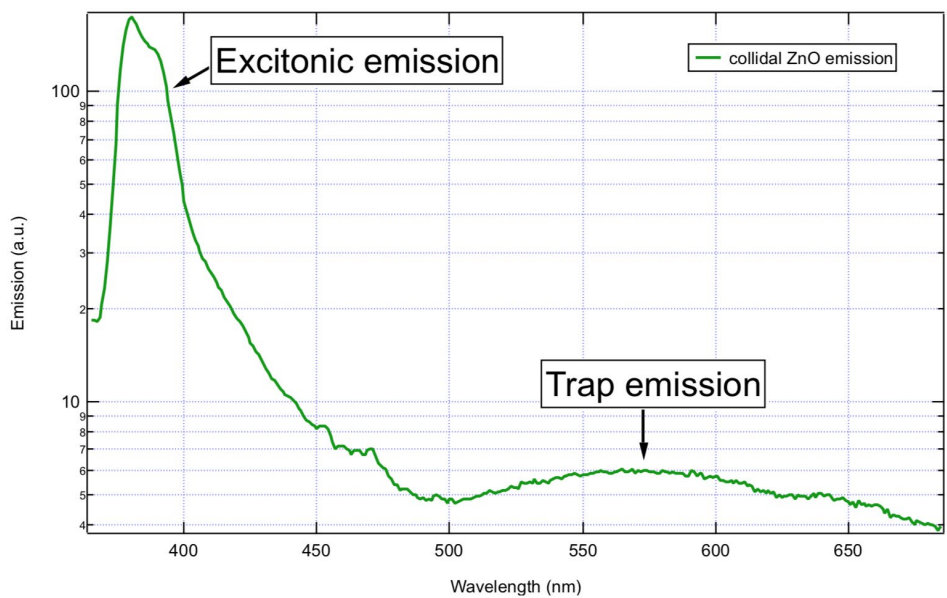

Figure S2: Steady-state PL spectrum of ZnO NPs. From ref. [2]

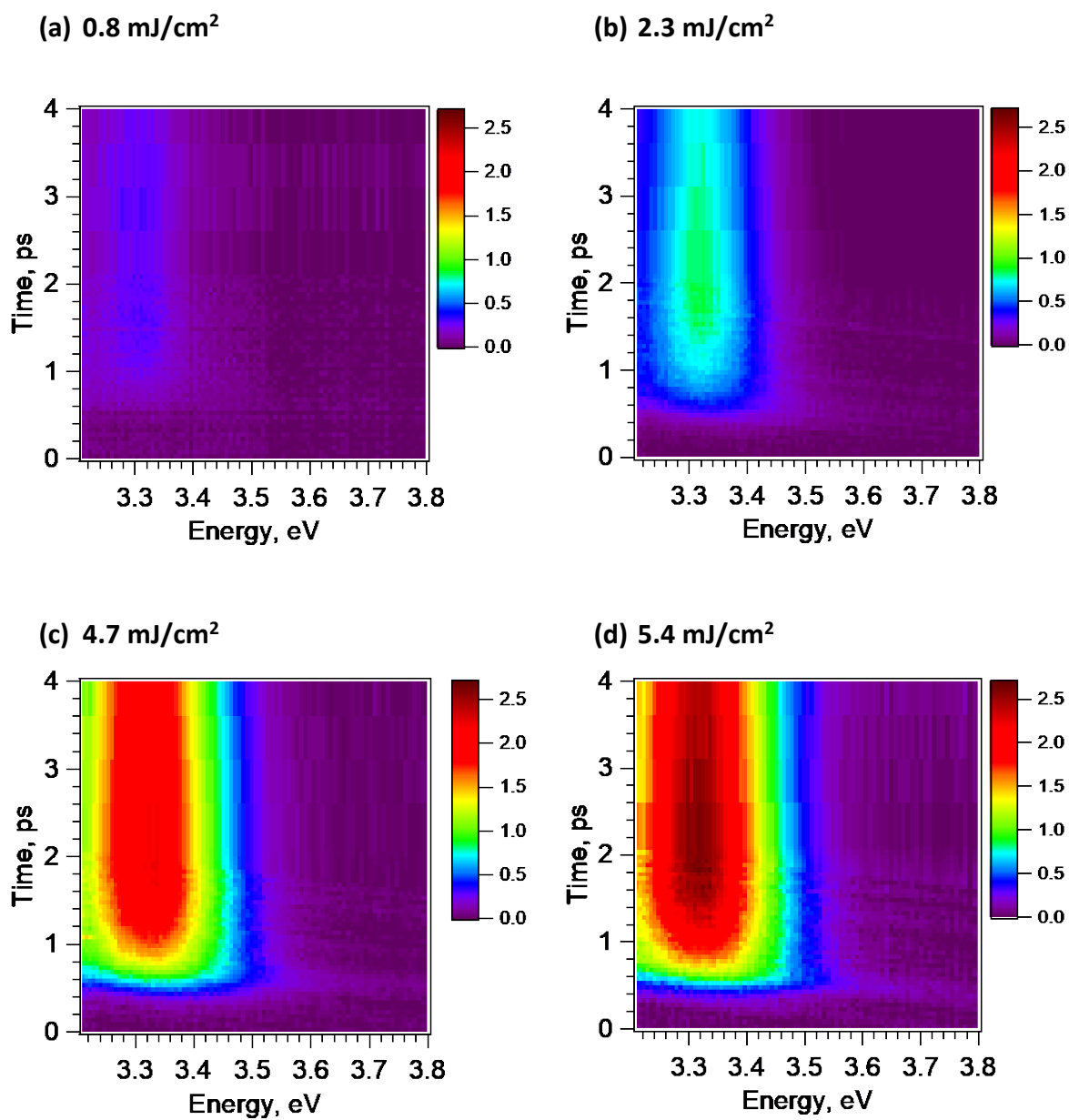

**Figure S3.** Time-energy plots of the PL of 266 nm-excited ZnO NPs measured at four different excitation beam fluences.

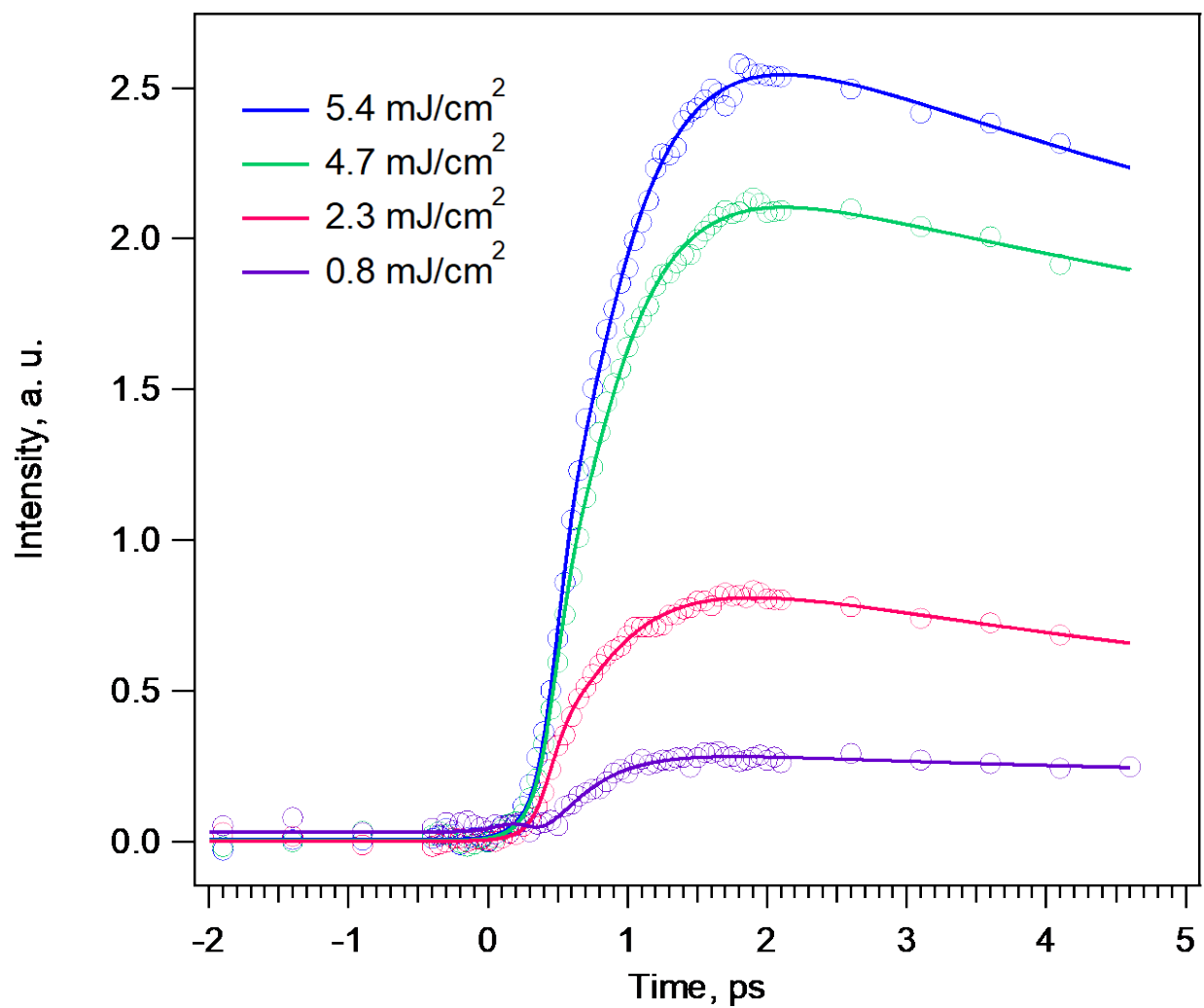

**Figure S4**. Kinetic traces at an energy of 3.31 eV from the time-energy emission data of the 4.66 eV-excited ZnO NPs (Figure S1) measured at four different fluence values of the excitation beam. Experimental data (open circles) are fitted by a multiexponential function convoluted with the instrumental response function (IRF) of the setup (solid lines). The obtained time constants are summarized in the Table S2.

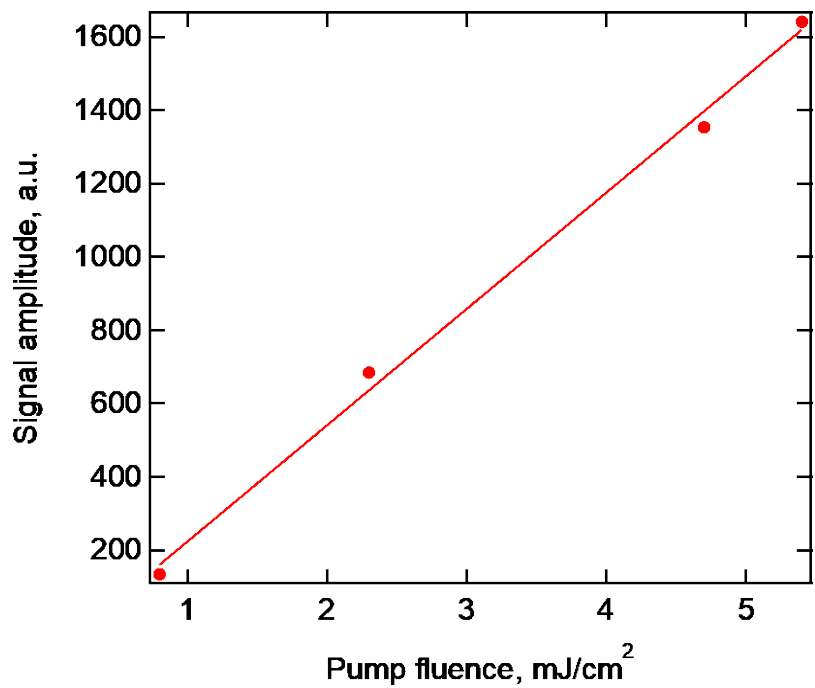

**Figure S5**. Dependence of the ZnO PL emission intensity recorded at 3.3 eV as a function of the fluence of the 4.66 eV excitation beam.

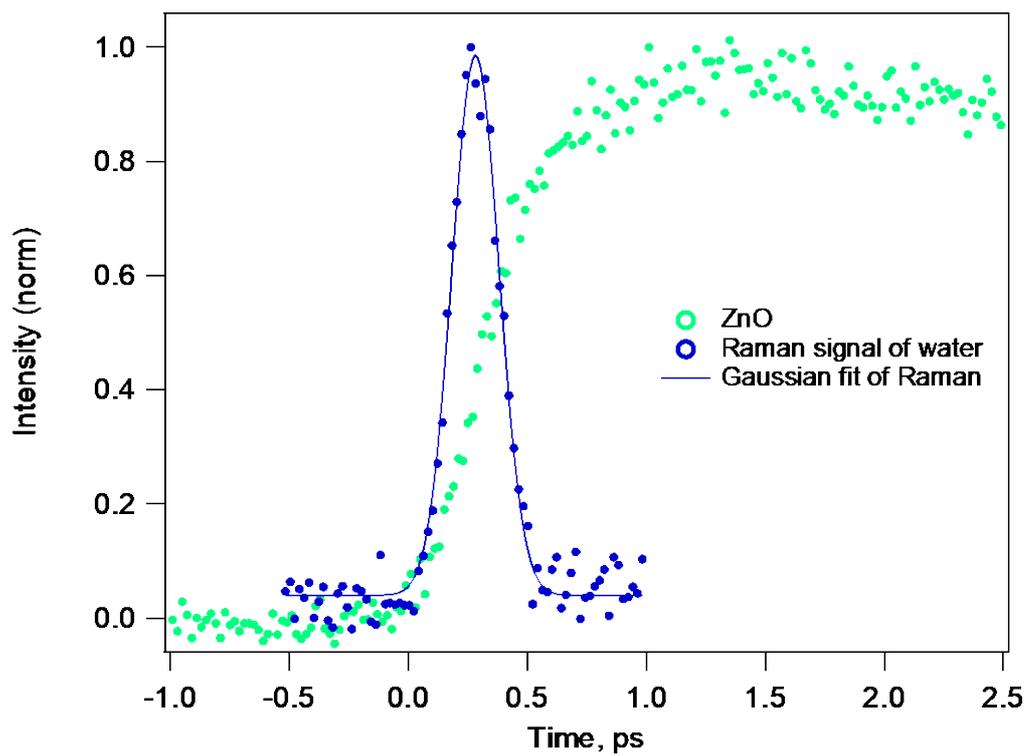

**Figure S6**. Time trace of the Raman signal of pure water (blue dots) recorded at 4.22 eV, and its Gaussian fit (blue line), which yields the Instrument response function (IRF) of 230 fs, and kinetic trace of the ZnO NPs' PL (green dots) detected at 3.31 eV. Both samples are excited with 266 nm.

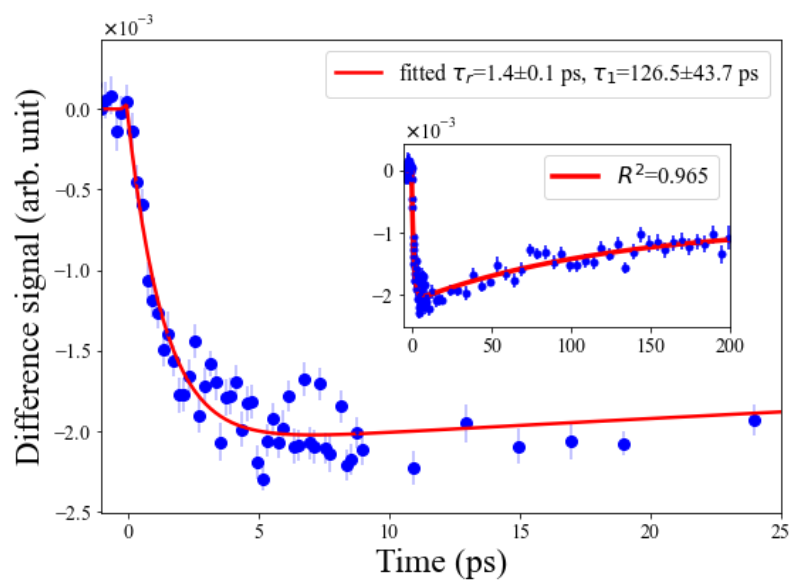

**Figure S7:** Fit of the fs-XANES time trace at 9.67 keV using a rise and a decay component. See § S.2 for details

[1] T. Rossi, T. J. Penfold, M. H. Rittmann-Frank, M. Reinhard, J. Rittmann, C. N. Borca, D. Grolimund, C. J. Milne, and M. Chergui, *Characterizing the Structure and Defect Concentration of ZnO Nanoparticles in a Colloidal Solution*, J. Phys. Chem. C **118**, 19422 (2014).

[2] T. J. Penfold et al., *Revealing Hole Trapping in Zinc Oxide Nanoparticles by Time-Resolved X-Ray Spectroscopy*, Nat. Commun. **9**, 1 (2018).

[3] A. Cannizzo, O. Bram, G. Zgrablic, A. Tortschanoff, A. A. Oskouei, F. van Mourik, and M. Chergui, *Femtosecond Fluorescence Upconversion Setup with Broadband Detection in the Ultraviolet*, Opt. Lett. **32**, 24 (2007).

[4] T. Sato, T. Togashi, K. Ogawa, T. Katayama, Y. Inubushi, K. Tono, and M. Yabashi, *Highly Efficient Arrival Timing Diagnostics for Femtosecond X-Ray and Optical Laser Pulses*, Appl. Phys. Express **8**, 012702 (2014).

[5] T. Katayama et al., *A Beam Branching Method for Timing and Spectral Characterization of Hard X-Ray Free-Electron Lasers*, Struct. Dyn. **3**, 3 (2016).

[6] T. Katayama et al., *A Beam Branching Method for Timing and Spectral Characterization of Hard X-Ray Free-Electron Lasers*, Struct. Dyn. **3**, 034301 (2016).

[7] T. Katayama, S. Nozawa, Y. Umena, S. Lee, T. Togashi, S. Owada, and M. Yabashi, *A Versatile Experimental System for Tracking Ultrafast Chemical Reactions with X-Ray Free-Electron Lasers*, Struct. Dyn. **6**, 054302 (2019).

[8] S. J. Clark, M. D. Segall, C. J. Pickard, P. J. Hasnip, M. I. J. Probert, K. Refson, and M. C. Payne, *First Principles Methods Using CASTEP*, Z. Für Krist. - Cryst. Mater. **220**, 567 (2005).

[9] J. P. Perdew, K. Burke, and M. Ernzerhof, *Generalized Gradient Approximation Made Simple*, Phys. Rev. Lett. **77**, 3865 (1996).